\documentstyle[multicol,eqsecnum,aps,psfig]{revtex}
\renewcommand{\narrowtext}{\begin{multicols}{2}
\global\columnwidth20.5pc\noindent}
\renewcommand{\widetext}{\end{multicols}
\global\columnwidth42.5pc}
\begin{document}
\draft
\preprint{2 September 2003}
\title{Bosonic representation of one-dimensional Heisenberg ferrimagnets}
\author{Shoji Yamamoto}
\address{Division of Physics, Hokkaido University,
         Sapporo 060-0810, Japan}
\date{2 September 2003}
\maketitle
\begin{abstract}
The energy structure and the thermodynamics of ferrimagnetic Heisenberg
chains of alternating spins $S$ and $s$ are described in terms of the
Schwinger bosons and modified spin waves.
In the Schwinger representation, we average the local constraints on the
bosons and diagonalize the Hamiltonian at the Hartree-Fock level.
In the Holstein-Primakoff representation, we optimize the free energy
in two different ways introducing an additional constraint on the staggered
magnetization.
A new modified spin-wave scheme, which employs a Lagrange multiplier
keeping the native energy structure free from temperature and thus differs
from the original Takahashi Scheme, is particularly stressed as an
excellent language to interpret one-dimensional quantum ferrimagnetism.
Other types of one-dimensional ferrimagnets and the antiferromagnetic
limit $S=s$ are also mentioned.
\end{abstract}
\narrowtext

\section{Introduction}

   Significant efforts have been devoted to synthesizing low-dimensional
ferrimagnets and understanding their quantum behavior in recent years.
The first example of one-dimensional ferrimagnets,
MnCu(S$_2$C$_2$O$_2$)$_2$(H$_2$O)$_3$$\cdot$4.5H$_2$O, was synthesized by
Gleizes and Verdaguer \cite{G7373} and followed by a series of ordered
bimetallic chain compounds \cite{K89} in an attempt to design
molecule-based ferromagnets \cite{OK782}.
Caneschi {\it et al.} \cite{C1756} demonstrated another approach to
alternating-spin chains hybridizing manganese complexes and nitronyl
nitroxide radicals.
The inorganic-organic hybrid strategy realized more complicated alignments
of mixed spins \cite{I219}.
There also exists an attempt at stacking novel triradicals into a purely
organic ferrimagnet \cite{H7921}. 
Mono-spin chains can be ferrimagnetic with polymerized exchange
interactions.
An example of such ferrimagnets is the ferromagnetic-antiferromagnetic
bond-alternating copper tetramer chain compound
Cu(C$_5$H$_4$NCl)$_2$(N$_3$)$_2$ \cite{E4466}.
The trimeric intertwining double-chain compound
Ca$_3$Cu$_3$(PO$_4$)$_4$ \cite{D83}
is another solution to homometallic one-dimensional ferrimagnets, where
the noncompensation of sublattice magnetizations is of topological origin.
Besides one-dimensional ferrimagnets, metal-ion magnetic clusters such as
[Mn$_{12}$O$_{12}$(CH$_3$COO)$_{16}$(H$_2$O)$_4$] \cite{L2042} and
[Fe$_8$(N$_3$C$_6$H$_{15}$)$_6$O$_2$(OH)$_{12}$]$^{8+}$ \cite{W77},
for which resonant magnetization tunneling \cite{F3830,T145,S4645,W133}
was observed, are also worth mentioning as zero-dimensional ferrimagnets.

   The discovery of ordered bimetallic chain compounds stimulated
extensive theoretical interest in (quasi-)one-dimensional quantum
ferrimagnets.
Early efforts \cite{D413} were devoted to numerically diagonalizing
alternating-spin Heisenberg chains.
Numerical diagonalization, combined with the Lanczos algorithm
\cite{Y3711,A144414} and a scaling technique \cite{S4053}, further
contributed to studying modern topics such as phase transitions of the
Kosterlitz-Thouless type \cite{A144414,Y5175} and quantized magnetization
plateaux \cite{K1762,Y3795}.
Alternating-spin chains were further investigated by density-matrix
renormalization-group \cite{P8894,M5908} and quantum Monte Carlo
\cite{Y13610,Y1024} methods in an attempt to illuminate dual features of
ferrimagnetic excitations.
More general mixed-spin chains were analyzed via the nonlinear $\sigma$
model \cite{F14709} with particular emphasis on the competition between
massive and massless phases.
Quasi-one-dimensional mixed-spin systems \cite{M68,T15189} were also
investigated in order to explain the inelastic-neutron-scattering findings
\cite{Z385,Z7210} for the rare-earth nickelates $R_2$BaNiO$_5$.

   In order to complement numerical tools and to achieve further
understanding of the magnetic double structure of ferrimagnetism, several
authors have recently begun to construct bosonic theories of
low-dimensional quantum ferrimagnets.
The conventional spin-wave description of the ground-state properties
\cite{P8894,B3921,I14024,I144429}, a modified spin-wave scheme for the
low-temperature properties \cite{O8067}, and the Schwinger-boson
representation of the low-energy structure \cite{C915} and the
thermodynamics \cite{W1057}, they all reveal the potential of bosonic
languages for various ferrimagnetic systems.
However, considering the global argument and total understanding over
the bosonic theory of ferromagnets and antiferromagnets
\cite{T168,T2494,H4769,R2589,Y769,I1082,K104427,A316,H2850,S5028,L129},
ferrimagnets are still undeveloped in this context especially in one
dimension.
In such circumstances, 
we represent one-dimensional Heisenberg ferrimagnets in terms of the
Schwinger bosons and the Holstein-Primakoff spin waves.
Based on a mean-field ansatz, the local constraints on the Schwinger
bosons are relaxed and imposed only on the average.
The conventional antiferromagnetic spin-wave formalism \cite{A694,K568}
is modified, on the one hand following the Takahashi scheme
\cite{T168,T2494} which was originally proposed for ferromagnets, while
on the other hand introducing a slightly different new strategy
\cite{Y14008}.
The Schwinger bosons and the modified spin waves both interpret the
low-energy properties fairly well identifying the ferrimagnetic long-range
order with a Bose condensation, while the two languages are qualitatively
distinguished in describing the thermodynamics.
We demonstrate that {\it the new modified spin-wave scheme of all others
depict one-dimensional ferrimagnetic features very well}.

\section{Formalism}

   A practical model for one-dimensional ferrimagnets is two kinds of
spins, $S$ and $s$ ($S>s$), alternating on a ring with antiferromagnetic
exchange coupling between nearest neighbors, as described by the
Hamiltonian,
\begin{equation}
   {\cal H}
      =J\sum_{n=1}^N
        \left(
         \mbox{\boldmath$S$}_{n} \cdot \mbox{\boldmath$s$}_{n-1}
        +\mbox{\boldmath$s$}_{n} \cdot \mbox{\boldmath$S$}_{n}
        \right)\,,
   \label{E:H}
\end{equation}
where $N$ is the number of unit cells.
The simplest case, $(S,s)=(1,\frac{1}{2})$, has so far been discussed
fairly well using the matrix-product formalism \cite{K3336}, a modified
spin-wave scheme \cite{Y14008}, the Schwinger-boson representation
\cite{W1057}, and modern numerical techniques
\cite{P8894,M5908,Y13610,Y1024}.
We make further explorations into higher-spin systems and develop the
analytic argument in more detail.

\subsection{Schwinger-boson mean-field theory}

   Let us describe each spin variable in terms of two kinds of bosons as
\begin{equation}
   \begin{array}{ll}
   S_n^+=a_{n\uparrow}^\dagger a_{n\downarrow}\,,&
   S_n^z=\frac{1}{2}
         \bigl(a_{n\uparrow}^\dagger a_{n\uparrow}
              -a_{n\downarrow}^\dagger a_{n\downarrow}\bigr),\\
   s_n^+=b_{n\uparrow}^\dagger b_{n\downarrow}\,,&
   s_n^z=\frac{1}{2}
         \bigl(b_{n\uparrow}^\dagger b_{n\uparrow}
              -b_{n\downarrow}^\dagger b_{n\downarrow}\bigr),\\
   \label{E:SB}
   \end{array}
\end{equation}
where the constraints
\begin{equation}
   \sum_{\sigma=\uparrow,\downarrow}
    a_{n\sigma}^\dagger a_{n\sigma}
   =2S\,,\ \ 
   \sum_{\sigma=\uparrow,\downarrow}
    b_{n\sigma}^\dagger b_{n\sigma}
   =2s\,,
   \label{E:constSBL}
\end{equation}
are imposed on the bosons.
Then the Hamiltonian can be written as
\begin{equation}
   {\cal H}
   =2NJSs
   -2J\sum_{n=1}^N
    \left(
     \Omega_{n+}^\dagger\Omega_{n+}
    +\Omega_{n-}^\dagger\Omega_{n-}
    \right),
   \label{E:HSB}
\end{equation}
where
$\Omega_{n+}
 =(a_{n\uparrow}b_{n\downarrow}-a_{n\downarrow}b_{n\uparrow})/2$ and
$\Omega_{n-}
 =(a_{n\uparrow}b_{n-1\downarrow}-a_{n-1\downarrow}b_{n\uparrow})/2$.
The Hartree-Fock treatment assumes the thermal average of the short-range
antiferromagnetic order to be uniform and static as
\begin{equation}
   \langle\Omega_{n\pm}^\dagger\rangle_T
  =\langle\Omega_{n\pm}\rangle_T
  =\Omega\,.
\end{equation}
The constraints (\ref{E:constSBL}) are correspondingly relaxed as
\begin{equation}
   \sum_{n=1}^N\sum_{\sigma=\uparrow,\downarrow}
    a_{n\sigma}^\dagger a_{n\sigma}
   =2NS,\ 
   \sum_{n=1}^N\sum_{\sigma=\uparrow,\downarrow}
    b_{n\sigma}^\dagger b_{n\sigma}
   =2Ns.
   \label{E:constSBNL}
\end{equation}
In the momentum space the mean-field Hamiltonian reads
\begin{eqnarray}
   &&
   {\cal H}_{\rm MF}
   =2NJSs+4NJ\Omega^2-4NJ(\lambda S+\mu s)
   \nonumber\\
   &&\qquad\quad
   -2J\Omega\sum_k
    \cos ak
    \left(
     a_{k\uparrow}b_{k\downarrow}-a_{k\downarrow}b_{k\uparrow}
    +{\rm H.c.}
    \right)
   \nonumber\\
   &&\qquad\quad
   +2J\sum_k\sum_{\sigma=\uparrow,\downarrow}
    \left(
     \lambda a_{k\sigma}^\dagger a_{k\sigma}
    +\mu b_{k\sigma}^\dagger b_{k\sigma}
    \right),
   \label{E:HSBMF}
\end{eqnarray}
where $k$ is defined as $n\pi/Na$ $(n=0,1,\cdots,N-1)$ with $a$ being the
distance between neighboring spins, and $\lambda$ and $\mu$ are the
Lagrange multipliers due to the constraints (\ref{E:constSBNL}).
Via the Bogoliubov transformation
\begin{equation}
   \begin{array}{lll}
     a_{k\uparrow}
     &=&
     \alpha_{k\uparrow}{\rm cosh}\theta_k
    +\beta_{k\downarrow}^\dagger{\rm sinh}\theta_k\,,\\
     a_{k\downarrow}
     &=&
     \alpha_{k\downarrow}{\rm cosh}\theta_k
    -\beta_{k\uparrow}^\dagger{\rm sinh}\theta_k\,,\\
     b_{k\uparrow}
     &=&
     \beta_{k\uparrow}{\rm cosh}\theta_k
    -\alpha_{k\downarrow}^\dagger{\rm sinh}\theta_k\,,\\
     b_{k\downarrow}
     &=&
     \beta_{k\downarrow}{\rm cosh}\theta_k
    +\alpha_{k\uparrow}^\dagger{\rm sinh}\theta_k\,,
   \end{array}
   \label{E:BTSB}
\end{equation}
with
\begin{equation}
   {\rm tanh}2\theta_k=\frac{2\Omega\cos ak}{\lambda+\mu}\,,
   \label{E:thetaSB}
\end{equation}
the Hamiltonian (\ref{E:HSBMF}) is diagonalized as
\begin{eqnarray}
   &&
   {\cal H}_{\rm MF}
   =2NJSs+4NJ\Omega^2
   \nonumber\\
   &&\qquad\quad
   -2NJ\lambda(2S+1)-2NJ\mu(2s+1)
   +2J\sum_k\omega_k
   \nonumber\\
   &&\qquad\quad
   +J\sum_k\sum_{\sigma=\uparrow,\downarrow}
    \left(
     \omega_{k\sigma}^-\alpha_{k\sigma}^\dagger\alpha_{k\sigma}
    +\omega_{k\sigma}^+\beta_{k\sigma}^\dagger\beta_{k\sigma}
    \right),
\end{eqnarray}
where
\begin{eqnarray}
   &&
   \omega_{k\sigma}^{\pm}\equiv\omega_k^{\pm}
   =\omega_k\pm(\mu-\lambda);
   \nonumber\\
   &&
   \omega_k
   =\sqrt{(\lambda+\mu)^2-4\Omega^2\cos^2 ak}\,.
   \label{E:dspSB}
\end{eqnarray}
$\lambda$, $\mu$, and $\Omega$ are determined through a set of equations
\begin{eqnarray}
   &&
   \sum_k
   \left(
    \bar{n}_{k\sigma}^-{\rm cosh^2}\theta_k
   +\bar{n}_{k\sigma}^+{\rm sinh^2}\theta_k
   +{\rm sinh}^2\theta_k
   \right)=NS\,,
   \\
   &&
   \sum_k
   \left(
    \bar{n}_{k\sigma}^-{\rm sinh^2}\theta_k
   +\bar{n}_{k\sigma}^+{\rm cosh^2}\theta_k
   +{\rm sinh}^2\theta_k
   \right)=Ns\,,
   \\
   &&
   \sum_k
   \left(
    \bar{n}_{k\sigma}^- +\bar{n}_{k\sigma}^+ +1
   \right)
   {\rm cosh}\theta_k{\rm sinh}\theta_k=N\Omega\,,
\end{eqnarray}
where the thermal distribution functions
$\bar{n}_{k\sigma}^-
 \equiv\langle\alpha_{k\sigma}^\dagger\alpha_{k\sigma}\rangle_T$ and
$\bar{n}_{k\sigma}^+
 \equiv\langle\beta_{k\sigma}^\dagger\beta_{k\sigma}\rangle_T$
are required to minimize the free energy and given by
\begin{equation}
   \bar{n}_{k\sigma}^\pm
   =\frac{1}{{\rm e}^{\omega_{k\sigma}^\pm/k_{\rm B}T}-1}\,.
\end{equation}

   The magnetic susceptibility is expressed as
\begin{equation}
   \chi=\frac{(g\mu_{\rm B})^2}{4k_{\rm B}T}
        \sum_k\sum_{\tau=\pm}\sum_{\sigma=\uparrow,\downarrow}
        \bar{n}_{k\sigma}^\tau\left(\bar{n}_{k\sigma}^\tau+1\right),
   \label{E:chiSB}
\end{equation}
where we have set the $g$-factors of spins $\mbox{\boldmath$S$}$ and
$\mbox{\boldmath$s$}$ both equal to $g$.
The internal energy should be given by
\begin{equation}
   E=\frac{1}{2}(E_{\rm MF}+2NJSs)-2NJSs\,,
   \label{E:ESB}
\end{equation}
where
\begin{eqnarray}
   &&
   E_{\rm MF}
   =2NJSs+4NJ\Omega^2-2NJ(2\lambda S+2\mu s+\lambda+\mu)
   \nonumber\\
   &&\qquad\quad
   +2J\sum_k\omega_k
   +J\sum_k\sum_{\tau=\pm}\sum_{\sigma=\uparrow,\downarrow}
    \bar{n}_{k\sigma}^\tau\omega_{k\sigma}^\tau\,.
\end{eqnarray}
Arovas and Auerbach \cite{A316} pointed out that relaxing the original
constraints (\ref{E:constSBL}) into Eq. (\ref{E:constSBNL}) leads to
double counting the number of independent boson degrees of freedom.
Therefore, in Eq. (\ref{E:ESB}), we have corrected the mean-field artifact
reducing the overestimated quantum fluctuation.

\subsection{Modified spin-wave theory: Takahashi scheme}

   Next we consider a single-component bosonic representation of each spin
variable at the cost of the rotational symmetry.
We start from the Holstein-Primakoff transformation
\begin{equation}
   \left.
   \begin{array}{ll}
    S_n^+=\sqrt{2S-a_n^\dagger a_n}\ a_n\,,&
    S_n^z=S-a_n^\dagger a_n\,,\\
    s_n^+=b_n^\dagger\sqrt{2s-b_n^\dagger b_n}\,,&
    s_n^z=-s+b_n^\dagger b_n\,.
   \end{array}
   \right.
   \label{E:HP}
\end{equation}
Treating $S$ and $s$ as $O(S)=O(s)$, we can expand the Hamiltonian with
respect to $1/S$ as
\begin{equation}
   {\cal H}=-2NJSs+E_1+E_0+{\cal H}_1+{\cal H}_0+O(S^{-1}),
   \label{E:HHP}
\end{equation}
where $E_i$ and ${\cal H}_i$ give the $O(S^i)$ quantum corrections to the
ground-state energy and the dispersion relations, respectively.
Via the Bogoliubov transformation
\begin{equation}
   \left.
   \begin{array}{lll}
      a_k&=&
      \alpha_k{\rm cosh}\theta_k-\beta_k^\dagger{\rm sinh}\theta_k\,,\\
      b_k&=&
      \beta_k{\rm cosh}\theta_k-\alpha_k^\dagger{\rm sinh}\theta_k\,,
   \end{array}
   \right.
   \label{E:BTMSWT}
\end{equation}
they are written as
\begin{mathletters}
   \begin{eqnarray}
      E_1
      &=&-2NJ\left[2\sqrt{Ss}\,\Gamma-(S+s)\Lambda\right],\\
      E_0
      &=&-2NJ\Biggl[
              \Gamma^2+\Lambda^2
             -\Bigl(\sqrt{\frac{S}{s}}+\sqrt{\frac{s}{S}}\,\Bigr)
              \Gamma\Lambda
             \Biggr],
   \end{eqnarray}
\end{mathletters}
\begin{eqnarray}
   &&
   {\cal H}_i
   =J\sum_k
    \Bigl[
     \omega_i^-(k)\alpha_k^\dagger\alpha_k
    +\omega_i^+(k)\beta_k^\dagger\beta_k
   \nonumber\\
   &&\qquad\qquad
    +\gamma_i(k)
     \bigl(\alpha_k\beta_k+\alpha_k^\dagger\beta_k^\dagger\bigr)
    \Bigr],
   \label{E:HHPi}
\end{eqnarray}
where
\begin{eqnarray}
   &&
   \Gamma
    =\frac{1}{2N}\sum_k\cos ak\,{\rm sinh}2\theta_k\,,\\
   &&
   \Lambda
    =\frac{1}{2N}\sum_k({\rm cosh}2\theta_k-1),
\end{eqnarray}
\begin{mathletters}
   \begin{eqnarray}
   \omega_1^\pm(k)
    &=&(S+s){\rm cosh}2\theta_k-2\sqrt{Ss}\cos ak\,{\rm sinh}2\theta_k
    \nonumber\\
    &&
      \pm(S-s)\equiv\omega_k\pm(S-s),
   \label{E:dspLMSW}\\
   \omega_0^\pm(k)
    &=&\Biggl[
        \Bigl(\sqrt{\frac{S}{s}}+\sqrt{\frac{s}{S}}\,\Bigr)\Gamma
       -2\Lambda
       \Biggr]{\rm cosh}2\theta_k
    \nonumber\\
    &&
      -\Biggl[
        2\Gamma
       -\Bigl(\sqrt{\frac{S}{s}}+\sqrt{\frac{s}{S}}\,\Bigr)\Lambda
       \Biggr]\cos ak\,{\rm sinh}2\theta_k
    \nonumber\\
    &&\pm\Bigl(\sqrt{\frac{S}{s}}-\sqrt{\frac{s}{S}}\,\Bigr),
   \label{E:dspIMSW}
   \end{eqnarray}
   \label{E:dspMSW}
\end{mathletters}
\begin{mathletters}
   \begin{eqnarray}
   \gamma_1(k)
    &=&2\sqrt{Ss}\cos ak\,{\rm cosh}2\theta_k
      -(S+s){\rm sinh}2\theta_k\,,\\
   \gamma_0(k)
    &=&\Biggl[
        2\Gamma
       -\Bigl(\sqrt{\frac{S}{s}}+\sqrt{\frac{s}{S}}\,\Bigr)\Lambda
       \Biggr]\cos ak\,{\rm cosh}2\theta_k
    \nonumber\\
    &&
      -\Biggl[
        \Bigl(\sqrt{\frac{S}{s}}+\sqrt{\frac{s}{S}}\,\Bigr)\Gamma
       -2\Lambda
       \Biggr]{\rm sinh}2\theta_k\,.
   \end{eqnarray}
\end{mathletters}

   The conventional spin-wave scheme naively diagonalize the Hamiltonian
(\ref{E:HHP}) and ends up with the number of bosons diverging with
increasing temperature.
In order to suppress this thermal divergence, Takahashi \cite{T2494}
considered optimizing the bosonic distribution functions under zero
magnetization and obtained an excellent description of the low-temperature
thermodynamics for low-dimensional Heisenberg ferromagnets.
For ferrimagnets, this idea is still useful \cite{O8067,Y14008} but never
applies away from the low-temperature region as it is.
The zero-magnetization constraint plays a role of keeping the number of
bosons finite under ferromagnetic interactions but does not work so under
antiferromagnetic interactions.
Takahashi \cite{T2494} and Hirsch {\it et al.} \cite{H4769} proposed
constraining the staggered magnetization, instead of the uniform
magnetization, to be zero as the antiferromagnetic version of the
modified spin-wave theory.
Their scheme was applied to extensive antiferromagnets in both two
\cite{T2494,H4769,H2887,X6861,C7832,D13821} and one \cite{R2589,Y769}
dimensions.
The conventional spin-wave procedure assumes that spins on one sublattice
point predominantly up, while those on the other predominantly down.
The modified spin-wave treatment restores the sublattice symmetry.
We consider the naivest generalization of the antiferromagnetic modified
spin-wave scheme to ferrimagnets.

   The constraint of zero staggered magnetization reads
\begin{equation}
   \sum_n\left(a_n^\dagger a_n+b_n^\dagger b_n\right)=N(S+s).
   \label{E:constMSWT}
\end{equation}
In order to enforce this condition, we first introduce a Lagrange
multiplier and diagonalize the effective Hamiltonian
\begin{equation}
   \widetilde{\cal H}
   ={\cal H}
   +2J\nu\sum_n\left(a_n^\dagger a_n+b_n^\dagger b_n\right).
   \label{E:HMSWT}
\end{equation}
Then the ground-state energy and the dispersion relations are obtained as
\begin{eqnarray}
   &&
   E_{\rm g}=-2NJSs+\widetilde{E}_1\,;\ \ 
   \widetilde{E}_1=E_1+4NJ\Lambda\nu\,,
   \label{E:EgLMSWT}\\
   &&
   \omega_k^\pm=\widetilde{\omega}_1^\pm(k)\,;\ \ 
   \widetilde{\omega}_1^\pm(k)
    =\omega_1^\pm(k)+2\nu{\rm cosh}2\theta_k\,,
   \label{E:dspLMSWT}
\end{eqnarray}
keeping only the bilinear terms and as
\begin{eqnarray}
   &&
   E_{\rm g}=-2NJSs+\widetilde{E}_1+E_0\,,
   \label{E:EgIMSWT}\\
   &&
   \omega_k^\pm=\widetilde{\omega}_1^\pm(k)+\omega_0^\pm(k)\,,
   \label{E:dspIMSWT}
\end{eqnarray}
considering the $O(S^0)$ interactions as well.
In terms of the spin-wave distribution functions
\begin{equation}
   \bar{n}_k^\pm=\frac{1}{{\rm e}^{\omega_k^\pm/k_{\rm B}T}-1}\,,
\end{equation}
the internal energy and the magnetic susceptibility are expressed as
\cite{T233}
\begin{eqnarray}
   &&
   E=E_{\rm g}
    +\sum_k\sum_{\tau=\pm}\bar{n}_k^\tau\omega_k^\tau\,,
   \label{E:EMSW}\\
   &&
   \chi=\frac{(g\mu_{\rm B})^2}{3k_{\rm B}T}
        \sum_k\sum_{\tau=\pm}
        \bar{n}_k^\tau\left(\bar{n}_k^\tau+1\right).
   \label{E:chiMSW}
\end{eqnarray}
$\theta_k$, defining the Bogoliubov transformation (\ref{E:BTMSWT}),
is determined through
\begin{equation}
   \gamma_1(k)-2\nu{\rm sinh}2\theta_k
    \equiv\widetilde{\gamma}_1(k)=0\,,
   \label{E:thetaMSW}
\end{equation}
provided we treat ${\cal H}_0$ as a perturbation to ${\cal H}_1$.

\subsection{Modified spin-wave theory: A new scheme}

   Although the Takahashi scheme overcomes the difficulty of sublattice
magnetizations diverging thermally, the obtained thermodynamics is still
far from satisfactory (see Fig. \ref{F:SsC} later on).
Within the conventional spin-wave theory, the quantum spin reduction,
that is, the quantum fluctuation of the ground-state sublattice
magnetization per unit cell, reads
\begin{eqnarray}
   &&
   \langle a_n^\dagger a_n\rangle_{T=0}
  =\langle b_n^\dagger b_n\rangle_{T=0}\equiv\delta
   \nonumber\\
   &&\qquad
  =\int_0^\pi
   \frac{S+s}{\sqrt{(S-s)^2+4Ss\sin^2(k/2)}}
   \frac{{\rm d}k}{2\pi}
  -\frac{1}{2}\,,
  \label{E:delta}
\end{eqnarray}
and diverges at $S=s$.
The Takahashi scheme settles this {\it quantum divergence} as well as
the {\it thermal divergence}.
However, the number of bosons does not diverge in the ferrimagnetic ground
state.
Without quantum divergence, it is not necessary to modify the dispersion
relations (\ref{E:dspLMSW}) into the temperature-dependent form
(\ref{E:dspLMSWT}).
While the thermodynamics should be modified, the quantum mechanics may be
left as it is.

   Such an idea leads to the Bogoliubov transformation free from
temperature replacing Eq. (\ref{E:thetaMSW}) by $\gamma_1(k)=0$, that is,
\begin{equation}
   {\rm tanh}2\theta_k=\frac{2\sqrt{Ss}\cos ak}{S+s}\,.
   \label{E:thetaSW}
\end{equation}
The ground-state energy and the dispersion relations are simply given by
\begin{equation}
   E_{\rm g}=-2NJSs+E_1\,;\ \ 
   \omega_k^\pm=\omega_1^\pm(k),
\end{equation}
within the up-to-$O(S^1)$ treatment and by
\begin{equation}
   E_{\rm g}=-2NJSs+E_1+E_0\,;\ \ 
   \omega_k^\pm=\omega_1^\pm(k)+\omega_0^\pm(k),
\end{equation}
in the up-to-$O(S^0)$ treatment.
They are nothing but the $T=0$ findings in the Takahashi scheme.

   At finite temperatures we replace $\alpha_k^\dagger\alpha_k$ and
$\beta_k^\dagger\beta_k$ in the spin-wave Hamiltonian (\ref{E:HHPi}) by
\begin{equation}
   \bar{n}_k^\mp\equiv\sum_{n^-,n^+=0}^\infty n^\mp P_k(n^-,n^+),
\end{equation}
where $P_k(n^-,n^+)$ is the probability of $n^-$ ferromagnetic and $n^+$
antiferromagnetic spin waves appearing in the $k$-momentum state and
satisfies
\begin{equation}
   \sum_{n^-,n^+} P_k(n^-,n^+)=1\,,
   \label{E:constP}
\end{equation}
for all $k$'s.
Then the free energy is written as
\begin{eqnarray}
   &&
   F=E_{\rm g}
    +J\sum_k\sum_{n^-,n^+}P_k(n^-,n^+)
      \sum_{\tau=\pm}n^\tau\omega_k^\tau
   \nonumber\\
   &&\qquad
  +k_{\rm B}T\sum_k\sum_{n^-,n^+}
   P_k(n^-,n^+){\rm ln}P_k(n^-,n^+).
   \label{E:FMSWY}
\end{eqnarray}
We minimize the free energy with respect to $P_k(n^-,n^+)$ enforcing a
condition
\begin{eqnarray}
   &&
   \langle S_n^z-s_n^z\rangle_T+2\delta
   \equiv\langle:S_n^z-s_n^z:\rangle_T
   \nonumber\\
   &&\qquad
  =S+s-\frac{S+s}{N}\sum_k\sum_{\tau=\pm}\frac{\bar{n}_k^\tau}{\omega_k}
   =0\,,
   \label{E:constMSWY}
\end{eqnarray}
as well as the trivial constraints (\ref{E:constP}).
In the second-side compact expression, the normal ordering is taken with
respect to both operators $\alpha$ and $\beta$.
Equation (\ref{E:constMSWY}) claims that the thermal fluctuation
$(S+s)\sum_k(n_k^-+n_k^+)/\omega_k$ should cancel the {\it full}, or
{\it classical}, N\'eel order $(S+s)N$ rather than the {\it quantum
mechanically reduced} one $(S+s-2\delta)N$.
Without consideration of the quantum fluctuation $2\delta$, which is
absent from ferromagnets but peculiar to ferrimagnets, the present scheme
breaks even the conventional spin-wave achievement at low temperatures.
Numerically solving the thermodynamic Bethe-Ansatz equations, Takahashi
and Yamada \cite{T2808} suggested that the conventional spin-wave theory
correctly gives the low-temperature leading term of the specific heat.
Both the Takahashi scheme with Eq. (\ref{E:constMSWT}) and the new scheme
with Eq. (\ref{E:constMSWY}) indeed keep unchanged the conventional
spin-wave findings
\begin{equation}
   \frac{C}{Nk_{\rm B}}
   \sim\frac{3}{4}\sqrt{\frac{S-s}{Ss}}
       \frac{\zeta(\frac{3}{2})}{\sqrt{2\pi}}\,t^{1/2}\ \ 
       (T\rightarrow 0),
\end{equation}
where $t=k_{\rm B}T/J$ within the up-to-$O(S^1)$ treatment, while
$t=k_{\rm B}T/\gamma J$ with $\gamma=1+\Gamma/\sqrt{Ss}-(S+s)\Lambda/Ss$
in the up-to-$O(S^0)$ treatment.
The conventional spin-wave approach gives no quantitative information on
the magnetic susceptibility, whereas the modified theory reveals
\begin{equation}
   \frac{\chi J}{N(g\mu_{\rm B})^2}
   \sim\frac{Ss(S-s)^2}{3}\,t^{-2}\ \ 
       (T\rightarrow 0).
\end{equation}
In terms of the optimum distribution functions
\begin{equation}
   \bar{n}_k^\pm
   =\frac{1}{{\rm e}^{[J\omega_k^\pm-\nu(S+s)/\omega_k]/k_{\rm B}T}-1}\,,
\end{equation}
the free energy at the thermal equilibrium is written as
\begin{equation}
   F=E_{\rm g}+\nu(S+s)N
    -k_{\rm B}T\sum_k\sum_{\tau=\pm}
               {\rm ln}\left(1+\bar{n}_k^\tau\right),
\end{equation}
where $\nu$ is the Lagrange multiplier due to the constraint
(\ref{E:constMSWY}).

\section{Results}

\begin{figure}
\centerline
{\mbox{\psfig{figure=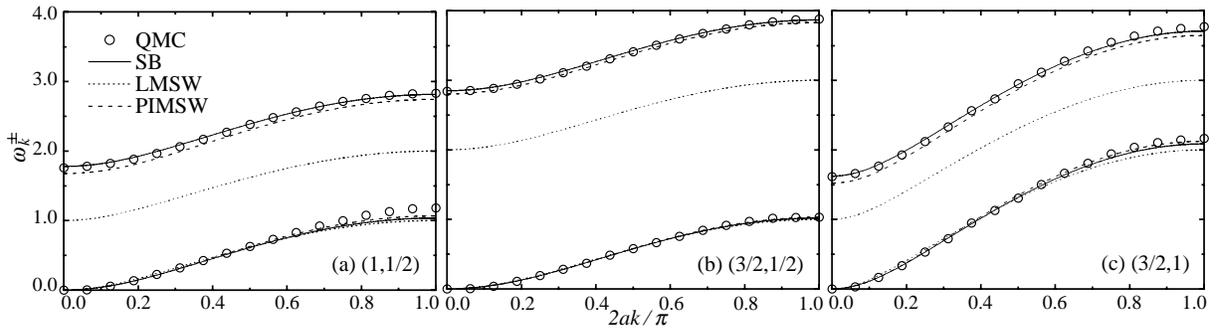,width=160mm,angle=0}}}
\vspace*{0.5mm}
\caption{The Schwinger-boson (SB), linear-modified-spin-wave (LMSW),
         perturbational interacting-modified-spin-wave (PIMSW), and
         quantum Monte Carlo (QMC) calculations of the dispersion
         relations of the ferromagnetic ($\omega_k^-$) and
         antiferromagnetic ($\omega_k^+$) elementary excitations for the
         spin-$(S,s)$ ferrimagnetic Heisenberg chains at zero
         temperature.}
\label{F:Ssdsp}
\end{figure}

   First we calculate the ground-state energy $E_g$ and the
antiferromagnetic excitation gap $\omega_{k=0}^+$ and compare them with
numerical findings in Table \ref{T:GSferri}.
At $T=0$, the Takahashi scheme and the new scheme lead to the same results
both giving $\nu=0$.
We are fully convinced that the spin-wave treatment better works for
larger spins.
We further learn that the spin-wave approach is better justified with
increasing $S/s$ as well as $Ss$, which is because the quantity $S-s$
fills the role of suppressing the divergence in Eq. (\ref{E:delta}).
On the other hand, the Schwinger-boson approach constantly gives highly
precise estimates of the low-energy properties.
Figure \ref{F:Ssdsp} further demonstrates that the Schwinger-boson
mean-field theory is highly successful in describing the low-lying
excitations.
Both the bosonic languages well interpret the ferromagnetic excitations,
whereas the linear spin waves considerably underestimate the
antiferromagnetic excitation energies.
The quantum correlation has much effect on the antiferromagnetic
excitation mode and such an effect is well included into the
Schwinger-boson calculation even at the mean-field level.

\begin{figure}
\centerline
{\mbox{\psfig{figure=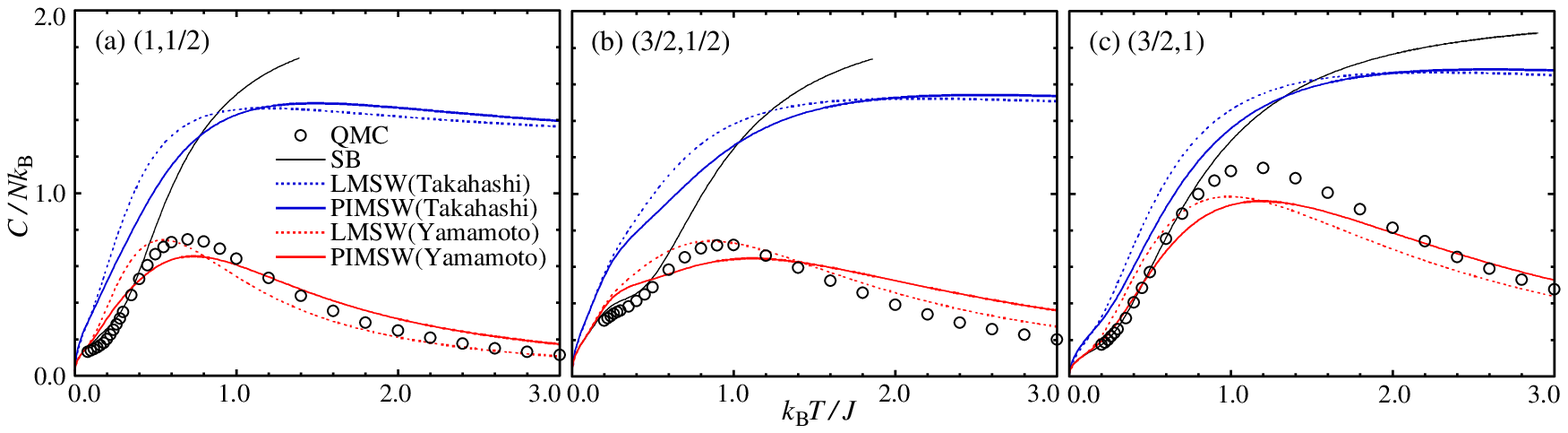,width=160mm,angle=0}}}
\vspace*{1mm}
\caption{The Schwinger-boson (SB), linear-modified-spin-wave (LMSW),
         perturbational interacting-modified-spin-wave (PIMSW), and
         quantum Monte Carlo (QMC) calculations of the specific heat $C$
         as a function of temperature for the spin-$(S,s)$ ferrimagnetic
         Heisenberg chains.
         The modified spin waves are constructed in two different ways,
         the Takahashi scheme (Takahashi) and the new scheme (Yamamoto).}
\label{F:SsC}
\end{figure}

   Next we calculate the thermodynamic properties.
Figure \ref{F:SsC} shows the temperature dependence of the specific heat.
The Schwinger-boson mean-field theory is still highly successful at low
temperatures, while with increasing temperature, it rapidly breaks down
failing to reproduce the Schottky-type peak.
The mean-field order parameter $\Omega$ monotonically decreases with
increasing temperature and reaches zero at
\begin{equation}
   \frac{k_{\rm B}T}{J}
   =\frac{S+s+1}{{\rm ln}(1+1/S)+{\rm ln}(1+1/s)}\,.
   \label{E:Tp}
\end{equation}
Above this temperature, $\Omega$ sticks at zero suggesting no
antiferromagnetic correlation in the system.
The onset of the paramagnetic phase at a finite temperature is a
mean-field artifact and the particular temperature (\ref{E:Tp}) is an
increasing function of $S$ and $s$.
The modified spin-wave theory based on the Takahashi scheme also fails to
describe the Schottky peak.
Because of the Lagrange multiplier $\nu$, which turns out a monotonically
increasing function of temperature, the dispersion relations
(\ref{E:dspLMSWT}) lead to endlessly increasing energy and thus
nonvanishing specific heat at high temperatures.
{\it Only the modified spin-wave theory based on the new scheme succeeds
in interpreting the Schottky peak}.
Since the antiferromagnetic excitation gap is significantly improved by
the inclusion of the $O(S^0)$ correlation, the interacting modified spin
waves reproduce the location of the Schottky peak fairly well.
Mixed-spin trimeric chain ferrimagnets have recently been synthesized
\cite{I219} and their low-temperature thermal properties were well
elucidated by the modified spin-wave theory \cite{O8067}.
However, it was unfortunate that the additional constraint was imposed on
the uniform magnetization and therefore the higer-temperature properties
were much less illuminated.
Controlling the staggered magnetization instead based on the new scheme,
we can fully investigate such polymeric chain compounds as well.
\widetext
\vspace*{-2.5mm}

\vspace*{-3mm}
\begin{figure}
\centerline
{\mbox{\psfig{figure=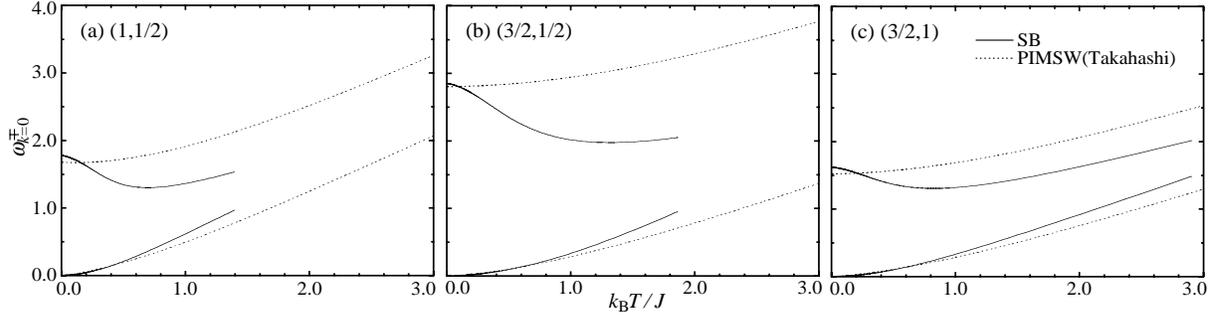,width=160mm,angle=0}}}
\vspace*{1mm}
\caption{The ferromagnetic ($\omega_{k=0}^-$) and antiferromagnetic
         ($\omega_{k=0}^+$) excitation gaps as functions of temperature
         for the spin-$(S,s)$ ferrimagnetic Heisenberg chains calculated
         by the Schwinger bosons (SB) and the perturbationally interacting
         modified spin waves (PIMSW) based on the Takahashi scheme.}
\label{F:Ssgap}
\end{figure}
\narrowtext

   In the Schwinger representation and the modified spin-wave treatment
based on the Takahashi scheme, the energy spectrum depends on temperature.
Since the low-energy band structure is well reflected in the thermal
behavior and can directly be observed through inelastic-neutron-scattering
measurements, we investigate the ferromagnetic ($\omega_{k=0}^-$) and
antiferromagnetic ($\omega_{k=0}^+$) excitation gaps as functions of
temperature in Fig. \ref{F:Ssgap}.
The Schwinger-boson mean-field theory claims that the antiferromagnetic
gap should first decrease and then increase with increasing temperature,
while the modified spin-wave theory predicts that the excitation energies
of both modes should be monotonically increasing functions of temperature.
We find a similar contrast between the two languages applied to ladder
ferrimagnets \cite{C915,N1380}.
In the case of Haldane-gap antiferromagnets, both the Schwinger-boson and
modified-spin-wave \cite{Y769} findings, together with the
nonlinear-$\sigma$-model calculations \cite{A474,J9265},
commonly suggest that the Haldane gap is a simply activated function of
temperature.
Extensive measurements on spin-$1$ antiferromagnetic Heisenberg chain
compounds \cite{R3538,R543,S3025} also report that the Haldane massive
mode is shifted upward with increasing temperature.
We encourage neutron-scattering experiments on ferrimagnetic chain
compounds to solve the present disagreement between the Schwinger-boson
and modified-spin-wave calculations of the antiferromagnetic excitation
gap as a function of temperature.

\begin{figure}
\centerline
{\mbox{\psfig{figure=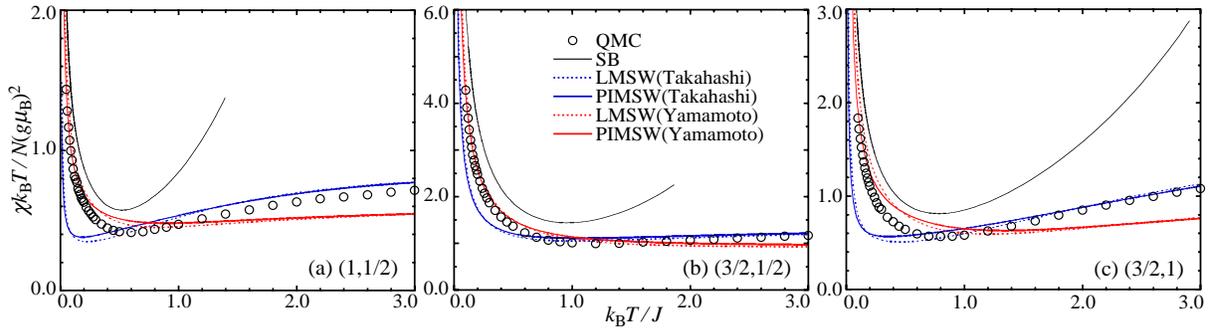,width=160mm,angle=0}}}
\vspace*{1mm}
\caption{The Schwinger-boson (SB), linear-modified-spin-wave (LMSW),
         perturbational interacting-modified-spin-wave (PIMSW), and
         quantum Monte Carlo (QMC) calculations of the
         susceptibility-temperature product $\chi T$ as a function of
         temperature for the spin-$(S,s)$ ferrimagnetic Heisenberg chains.
         The modified spin waves are constructed in two different ways,
         the Takahashi scheme (Takahashi) and the new scheme (Yamamoto).}
\label{F:SsST}
\end{figure}

   Figure \ref{F:SsST} shows the temperature dependence of the magnetic
susceptibility-temperature product, which elucidates ferromagnetic and
antiferromagnetic features coexisting in ferrimagnets \cite{Y1024}.
$\chi T$ diverges at low temperatures in a ferromagnetic fashion but
approaches the high-temperature paramagnetic behavior showing an
antiferromagnetic increase.
The modified spin waves much better describe the magnetic behavior than
the Schwinger bosons.
The spin waves modified along with the Takahashi scheme better work at
high temperatures, while those along with the new scheme precisely
reproduce the low-temperature behavior.
Both calculations converge into the paramagnetic behavior
$\chi k_{\rm B}T/N(g\mu_{\rm B})^2=[S(S+1)+s(s+1)]/3$ at high temperatures,
whereas the Schwinger-boson mean-field theory again breaks down at the
particular temperature (\ref{E:Tp}).
Considering that numerical tools less work at low temperatures, we realize
the superiority of the new-scheme-based modified spin-wave theory all the
more.

\begin{figure}
\centerline
{\mbox{\psfig{figure=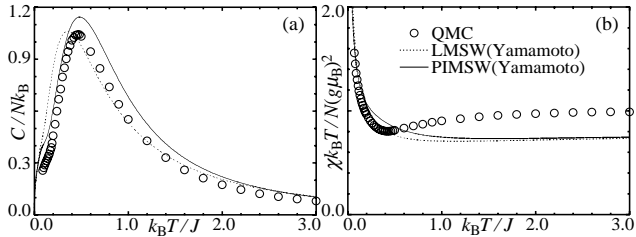,width=84mm,angle=0}}}
\vspace*{2mm}
\caption{The linear-modified-spin-wave (LMSW), perturbational
         interacting-modified-spin-wave (PIMSW), and quantum Monte Carlo
         (QMC) calculations of the specific heat $C$ and the 
         susceptibility-temperature product $\chi T$ as functions of
         temperature for the spin-$\frac{1}{2}$ bond-tetrameric
         ferrimagnetic Heisenberg chain of $J_{\rm F}=J_{\rm AF}$.
         The modified spin waves are constructed on the new scheme.}
\label{F:tetra}
\end{figure}

   Finally we calculate another type of ferrimagnet in order to
demonstrate the constant applicability of the present new scheme.
Figure \ref{F:tetra} shows the thermodynamic properties of the
ferromagnetic-ferromagnetic-antiferromagnetic-antiferromagnetic
bond-tetrameric spin-$\frac{1}{2}$ Heisenberg chain,
\widetext
\narrowtext
\begin{eqnarray}
   &&
   {\cal H}
   =\sum_{n=1}^N
    \big[
     J_{\rm AF}
     (\mbox{\boldmath$S$}_{4n-3}\cdot\mbox{\boldmath$S$}_{4n-2}
     +\mbox{\boldmath$S$}_{4n-2}\cdot\mbox{\boldmath$S$}_{4n-1})
   \nonumber \\
   &&\qquad\quad
    -J_{\rm F}
     (\mbox{\boldmath$S$}_{4n-1}\cdot\mbox{\boldmath$S$}_{4n  }
     +\mbox{\boldmath$S$}_{4n  }\cdot\mbox{\boldmath$S$}_{4n+1})
    \big]\,,
   \label{E:Htetra}
\end{eqnarray}
where we have set all the $g$-factors equal for simplicity.
The new modified spin-wave scheme again successfully reproduces the
Schottky peak of the Specific heat.
The interacting modified spin waves further interpret the low-temperature
Shoulder-like structure.
The characteristic minimum of the susceptibility-temperature product is
unfortunately less reproduced but the calculation again correctly gives
the paramagnetic susceptibility at sufficiently high temperatures.
A recent experiment \cite{H150} on a single-crystal sample of
Cu(C$_5$H$_4$NCl)$_2$(N$_3$)$_2$ \cite{E4466}, which may be described by
the Hamiltonian (\ref{E:Htetra}), has reported that the specific heat
exhibits a double-peaked structure as a function of temperature.
There is indeed a possibility of an additional peak appearing at low
temperatures as the ratio $J_{\rm F}/J_{\rm AF}$ moves away from unity
\cite{N214418}.
However, no parameter assignment has yet succeeded in interpreting all the
observations consistently.
There are further chemical attempts to synthesize novel ferrimagnets.
Organic ferrimagnets \cite{H7921,I2609} are free from magnetic anisotropy
and thus suitable for analyzing in terms of the modified spin waves.

\begin{figure}
\centerline
{\mbox{\psfig{figure=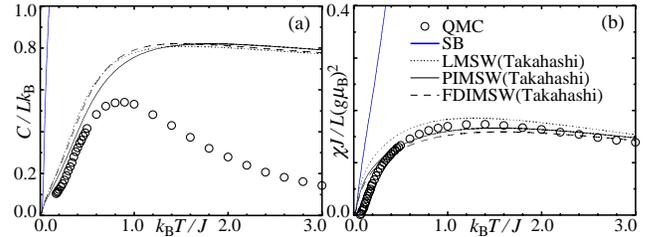,width=84mm,angle=0}}}
\vspace*{2mm}
\caption{The linear-modified-spin-wave (LMSW), perturbational
         interacting-modified-spin-wave (PIMSW), full-diagonalization
         interacting-modified-spin-wave (FDIMSW), and quantum Monte Carlo
         (QMC) calculations of the specific heat $C$ and the
         susceptibility $\chi$ as functions of temperature for the
         spin-$1$ antiferromagnetic Heisenberg chain.
         The modified spin waves are constructed on the Takahashi scheme.}
\label{F:Haldane}
\end{figure}

\section{Summary and Discussion}

   We have demonstrated the Schwinger-boson mean-field representation and
the modified spin-wave treatment of one-dimensional Heisenberg
ferrimagnets.
The Schwinger bosons form an excellent language at low temperatures but
rapidly lose their validity with increasing temperature.
The modified spin-wave theory is more reliable in totality provided the
number of bosons is controlled without modifying the native energy
structure.
On the other hand, the Schwinger-boson representation can be extended to
anisotropic systems \cite{L192} more reasonably because it is rotationally
invariant in contrast to the modified spin-wave theory.
While the temperature dependence of the anitiferromagnetic excitation gap
$\omega_{k=0}^+$ is left to solve experimentally, we are now convinced
that the bosonic languages remain effective in low dimensions and may be
applied to extensive ferrimagnets \cite{D3551}.
Besides ground-state properties and thermodynamics, quantum spin dynamics
\cite{Y157603,H054409} can be investigated through the modified spin-wave
scheme.

   We further mention our findings in the antiferromagnetic limit with the
view of realizing the close relation between the two bosonic languages.
We equalize $s$ with $S$ and set $2N$, the number of spins, equal to $L$
for the Hamiltonian (\ref{E:H}).
At $S=s$, the ground-state sublattice magnetization (\ref{E:delta})
diverges and therefore the new modified spin-wave scheme is no more
applicable.
We have to settle the quantum, as well as thermal, divergence inevitably
employing the Takahashi scheme.
Besides the perturbational treatment of ${\cal H}_0$, we may consider
the full diagonalization of ${\cal H}_1+{\cal H}_0$, where the
ground-state energy and the dispersion relations are still given by Eqs.
(\ref{E:EgIMSWT}) and (\ref{E:dspIMSWT}), respectively, but with
$\theta_k$ satisfying
\begin{equation}
   \widetilde{\gamma}_1(k)+\gamma_0(k)=0\,.
\end{equation}
Such an idea applied to ferrimagnets ends in gapped ferromagnetic
excitations and misreads the low-energy physics.
The perturbational series-expansion approach is highly successful in the
case of ferrimagnets \cite{I14024,I144429}.
Focussing our interest on Haldane-gap antiferromagnets, we list the
bosonic calculations of the ground-state properties in Table
\ref{T:GSHaldane}.
The bosonic languages interpret the ground-state correlation very well but
underestimate the Haldane gap considerably.
Indeed they cannot detect the topological terms responsible for vanishing
gap \cite{H464}, but they are still qualitatively consistent with the
nonlinear-$\sigma$-model quantum field theory, yielding the
low-temperature limiting behavior
$\omega_{k=0}^+-\Delta_0\propto{\rm e}^{-\Delta_0/T}$ \cite{Y769,J9265}
and the large-spin asymptotic behavior
$\Delta_0\propto{\rm e}^{-\pi S}$ \cite{A316,S5028,H464}.
The Schwinger-boson mean-field theory and the full-diagonalization
interacting modified spin-wave treatment give the same estimate of the
Haldane gap.
The Schwinger-boson dispersion relation (\ref{E:dspSB}) indeed coincides
analytically with that of the full-diagonalization interacting modified
spin waves at zero temperature.
This is interesting but not so surprising, because the Holstein-Primakoff
bosons (\ref{E:HP}) are obtained by replacing both
$a_{n\uparrow}$ ($b_{n\downarrow}$) and
$a_{n\uparrow}^\dagger$ ($b_{n\downarrow}^\dagger$) by
$\sqrt{2S-a_{n\downarrow}^\dagger a_{n\downarrow}}$
($\sqrt{2s-b_{n\uparrow}^\dagger b_{n\uparrow}}\,$) in the
transformation (\ref{E:SB}).

   Figure \ref{F:Haldane} shows the thermodynamic calculations for the
spin-$1$ antiferromagnetic Heisenberg chain.
We learn that the Schwinger-boson mean-field theory does not work at all
for spin-gapped antiferromagnets at finite temperatures, which is in
contrast with its fairly good representation of the low-temperature
thermodynamics for ferrimagnetic chains.
On the other hand, the modified spin-wave treatment maintains its validity
to a certain extent.
Indeed the Takahashi scheme still fails to reproduce the antiferromagnetic
Schottky-type peak of the specific heat, but it describes the
susceptibility very well except for the low-temperature findings
attributable to the underestimate of the Haldane gap.
We may expect the modified spin waves to efficiently depict the dynamic,
as well as static, susceptibility for extensive spin-gapped
antiferromagnets including spin ladders \cite{H}.
As for the thermal properties of one-dimensional antiferromagnets, whether
spin gapped or not, there is a possibility of a fermionic language
\cite{D964,H1607}, which is in principle compact, being superior to any
bosonic representation.

   In the case of ferromagnets, the Holstein-Primakoff bosons are already
diagonal in the momentum space \cite{T168,T233}, suggesting no quantum
fluctuation in the ground state, and therefore the present new scheme
turns out equivalent to the Takahashi scheme.
The new-scheme-based modified spin-wave theory is the very method for
low-dimensional ferrimagnets and is ready for extensive explorations.

\acknowledgments

   The authors are grateful to M. Takahashi and H. Hori for useful
discussions.
This work was supported by the Ministry of Education, Culture, Sports,
Science, and Technology of Japan, and the Nissan Science Foundation.

\widetext
\vskip 20mm
\begin{table}
\caption{The Schwinger-boson (SB), linear-modified-spin-wave (LMSW),
         perturbational interacting-modified-spin-wave (PIMSW), and
         numerical diagonalization (Exact) calculations of the
         ground-state energy $E_{\rm g}$ and the zero-temperature
         antiferromagnetic excitation gap $\Delta_0$ for the spin-$(S,s)$
         ferrimagnetic Heisenberg chains.}
\begin{tabular}{lllllll}
&
\multicolumn{2}{c}{$(S,s)=(1,\frac{1}{2})$} &
\multicolumn{2}{c}{$(S,s)=(\frac{3}{2},\frac{1}{2})$} &
\multicolumn{2}{c}{$(S,s)=(\frac{3}{2},1)$} \\
\cline{2-3}
\cline{4-5}
\cline{6-7}
{\raisebox{1.5ex}[0pt]{Approach}} &
$\ \ E_{\rm g}/NJ$ & $\ \ \Delta_0/J$ &
$\ \ E_{\rm g}/NJ$ & $\ \ \Delta_0/J$ &
$\ \ E_{\rm g}/NJ$ & $\ \ \Delta_0/J$ \\
\hline
SB    & $-1.45525$ & $1.77804$
      & $-1.96755$ & $2.84973$
      & $-3.86270$ & $1.62152$ \\
LMSW  & $-1.43646$ & $1      $
      & $-1.95804$ & $2      $
      & $-3.82807$ & $1      $ \\
PIMSW & $-1.46084$ & $1.67556$
      & $-1.96983$ & $2.80253$
      & $-3.86758$ & $1.52139$ \\
Exact & $-1.4541(1)$ & $1.759(1)$
      & $-1.9672(1)$ & $2.842(1)$
      & $-3.861(1) $ & $1.615(5)$ \\
\end{tabular}
\label{T:GSferri}
\end{table}

\begin{table}
\caption{The Schwinger-boson (SB), linear-modified-spin-wave (LMSW),
         perturbational interacting-modified-spin-wave (PIMSW),
         full-diagonalization interacting-modified-spin-wave (FDIMSW), and
         quantum Monte Carlo (QMC) [71] calculations of the
         ground-state energy $E_{\rm g}$ and the lowest excitation gap
         $\Delta_0$ for the spin-$S$ antiferromagnetic Heisenberg chains.}
\begin{tabular}{lllllll}
&
\multicolumn{2}{c}{$S=1$} &
\multicolumn{2}{c}{$S=2$} &
\multicolumn{2}{c}{$S=3$} \\
\cline{2-3}
\cline{4-5}
\cline{6-7}
{\raisebox{1.5ex}[0pt]{Approach}} &
$\ \ E_{\rm g}/LJ$ & $\ \ \Delta_0/J$ &
$\ \ E_{\rm g}/LJ$ & $\ \ \Delta_0/J$ &
$\ \ E_{\rm g}/LJ$ & $\ \ \Delta_0/J$ \\
\hline
SB     & $- 1.396148$ & $0.08507$
       & $- 4.759769$ & $0.00684$
       & $-10.1231  $ & $0.00295$ \\
LMSW   & $- 1.361879$ & $0.07200$
       & $- 4.726749$ & $0.00626$
       & $-10.0901  $ & $0.00279$ \\
PIMSW  & $- 1.394853$ & $0.07853$
       & $- 4.759760$ & $0.00655$
       & $-10.1231  $ & $0.00287$ \\
FDIMSW & $- 1.394617$ & $0.08507$
       & $- 4.759759$ & $0.00684$
       & $-10.1231  $ & $0.00295$ \\
QMC    & $- 1.401481(4)$ & $0.41048(6)$
       & $- 4.761249(6)$ & $0.08917(4)$
       & $-10.1239(1)  $ & $0.01002(3)$ \\
\end{tabular}
\label{T:GSHaldane}
\end{table}

\end{document}